\documentclass[12pt]{iopart}
\usepackage{iopams}
\usepackage{epsfig}
\usepackage{amssymb}
\usepackage{amsfonts}
\usepackage{mathrsfs}
\usepackage{epsfig}
\usepackage{bm}
\begin{document}

\title[Self-modulation instability of ultra-relativistic particle bunches with finite rise times]{Self-modulation instability of ultra-relativistic particle bunches with finite rise times}

\author{J. Vieira$^{1,5}$, L.D. Amorim$^1$, Y. Fang$^2$, W.B.Mori$^3$, P. Muggli$^4$, L.O. Silva$^1$}
\address{$^1$~GoLP/Instituto de Plasmas e Fus\~{a}o Nuclear-Laborat\'orio Associado,  Instituto Superior T\'{e}cnico, Universidade de Lisboa, 1049-001 Lisboa, Portugal}
\address{$^2$~University of Southern California, Los Angeles, CA 90089}
\address{$^3$~Department of Physics and Astronomy, University of California, Los Angeles, CA 90095}
\address{$^4$~Max Planck Institute for Physics, Munich, Germany}
\ead{$^5$jorge.vieira@ist.utl.pt}

\begin{abstract}
We study the evolution of the self-modulation instability using bunches with 
finite 
rise times. 
Using particle-in-cell simulations we show that unlike long bunches with sharp rise times, there are large variations of the wake amplitudes and wake phase velocity 
when bunches with finite rise times are used. 
These results show that use of bunches with sharp rise times is important to seed the self-modulation instability 
and 
to ensure stable acceleration regimes.
\end{abstract}


\maketitle

\section{\label{sec:introduction}Introduction}

Plasma acceleration uses intense laser pulse (laser wakefield acceleration - LWFA~\cite{bib:tajima_prl_1979}) or electron bunches (plasma wakefield acceleration - PWFA~\cite{bib:chen_prl_1986}) to drive large amplitude plasma waves propagating at nearly the speed of light. Plasma waves support acceleration gradients on the order of $E_{\mathrm{accel}}\simeq0.96\sqrt{n_0 [\mathrm{cm}^{-3}]}$ that are ideally suited to accelerate particles to high energies in short distances. Current plasma acceleration experiments use drivers shorter than the plasma wavelength to excite plasma waves in the so called bubble or blowout regime~\cite{bib:blowout}. In this regime the radiation pressure of an intense laser or the electrostatic repulsive force of a short electron bunch can radially expel all plasma electrons from a volume on the order of a cubic plasma 
wavelength. 
Plasma acceleration experiments in the LWFA (PWFA) blowout regime lead to the acceleration of 1 (42) GeV electrons in 1 cm (85 cm)~\cite{bib:acceleration}.

The use of proton bunches as drivers for plasma acceleration was recently proposed~\cite{bib:pdpwfa}. With more than 100 kJ, protons available at the Large Hadron Collider - CERN are very attractive for plasma based acceleration because they carry much more energy (7 TeV 
with 
 $10^{11}$ protons per bunch) than lepton bunches that will be produced by future linear colliders (500 GeV, with $10^{10}$ leptons and 1.6 kJ per bunch at the ILC). Numerical simulations demonstrated that compressed LHC proton bunches ($\sigma_z\simeq100~\mu$m) could lead to the acceleration of electrons by 600 GeV in a single 600 m long stage in a scheme know as the Proton Driven Plasma Wakefield Accelerator - PDPWFA~\cite{bib:pdpwfa}.

These simulations operated in the non-linear suck-in regime~\cite{bib:suckin}. Proton bunches produced by the LHC, however, are more than 10 cm long, and the compression of these bunches to $100~\mu\mathrm{m}$ is technically very demanding. These bunches are then too long 
and their density is to low 
to drive strongly non-linear plasma waves. Nevertheless, 
since 
they are much longer than the plasma wavelength at typical plasma densities $n_0\simeq 10^{15}~\mathrm{cm}^{-3}$ 
, they 
are ideally suited to generate large amplitude wakefields through the self-modulation instability (SMI)~\cite{bib:smi}. A self-modulated plasma wakefield accelerator will be explored experimentally at CERN during the next 3-5 years in a project known as AWAKE~\cite{awake}.

The self-modulation instability (SMI) consists in transverse modulations of the bunch radius, and it is driven by the plasma transverse focusing force. Bunch particles propagating in regions of focusing/defocusing regions are pulled towards/pushed away from the axis therefore decreasing/increasing the bunch radius at the plasma wavelength $\lambda_p$. As this occurs, the wakefield amplitude grows both along the bunch and also for longer propagation distances. In turn, as the wakefields become larger, the rate at which radial modulations occur also increases. SMI is then a convective instability, where longer bunches encompassing a larger number of plasma wavelengths and longer propagation distances, result in higher growth rates. SMI saturation occurs when the bunch becomes fully modulated into a train of beamlets separated by $\lambda_p$. The resulting bunch train resonantly excites wakefields that grow secularly from the head to the tail of the bunch.

\begin{figure}
\centering\includegraphics[width=0.7\columnwidth]{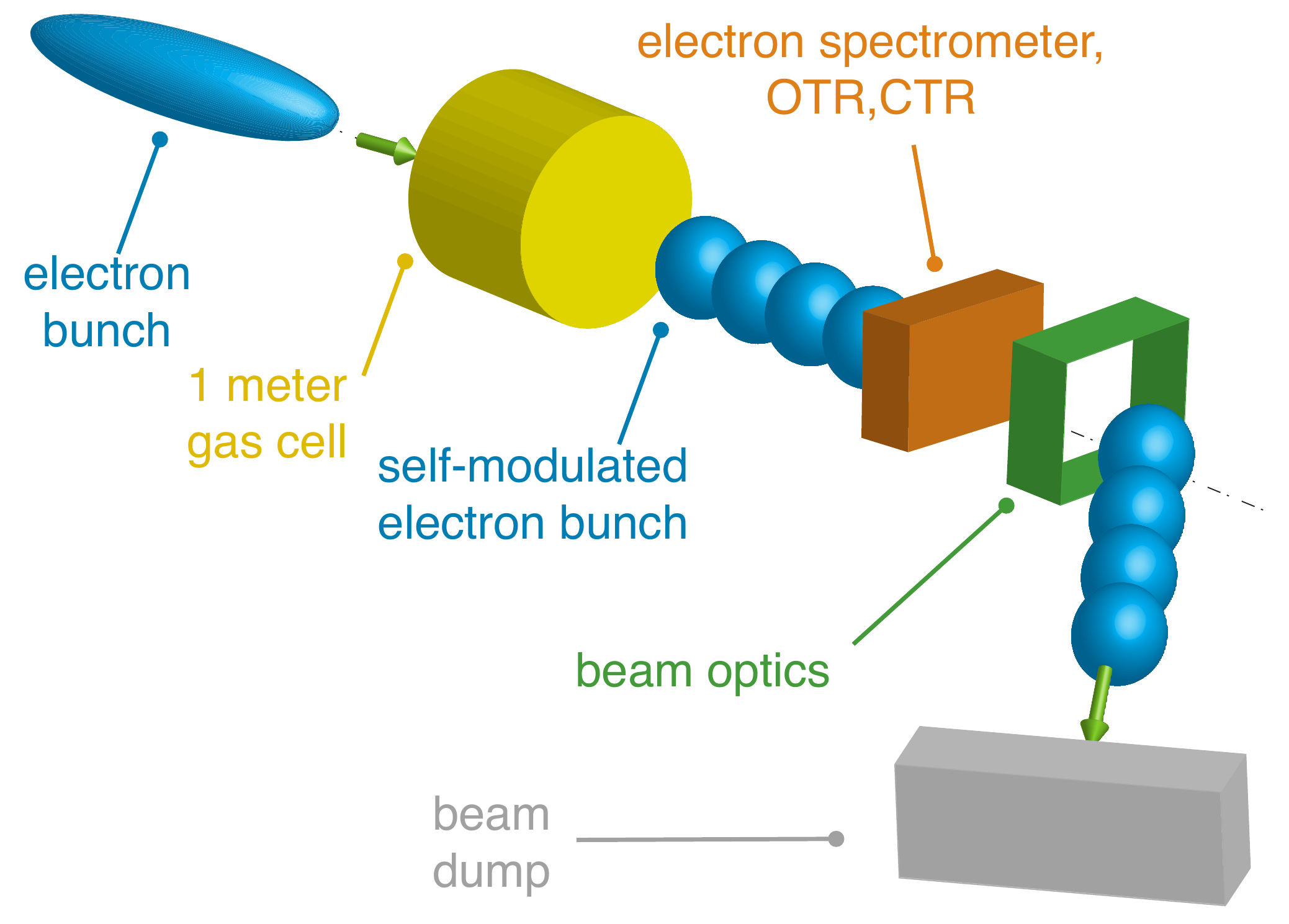}
\caption{Schematic representation of the experimental layout of a self-modulation experiment using uncompressed SLAC electron bunches. A long electron bunch with $\sigma_z \simeq 500-1000~\mathrm{\mu m}$ (blue ellipsoid) enters a meter long plasma cell (yellow cylinder) with densities $n_0\simeq 1\times10^{17}~\mathrm{cm}^{-3}$. The self-consistent interaction between the plasma and long electron bunch leading to the growth and saturation of the self-modulation results in a fully self-modulated long electron bunch (blue spheres). Bunch self-modulation can be measured through several diagnostics including Optical Transition Radiation (OTR) and Coherent Transition Radiation (CTR). The bunch energy after the plasma will also be measured by a magnetic spectrometer providing evidences of wakefield growth due to self-modulation.}
\label{fig:setup}
\end{figure}

Key SMI physics can be tested using electron and positron bunches available at SLAC FACET. This experiment will send an un-compressed 20 GeV SLAC FACET electron bunch with $2\times10^{10}$ particles into a plasma column with densities ranging from $n_0=10^{16}$ to $10^{17}~\mathrm{cm}^{-3}$. These un-compressed bunches have FWHM lengths varying between $500$ and $1500~
\mu\mathrm{m}$. %
Since at the above plasma densities the plasma wavelength varies between $\lambda_p = 100$ and $334~\mathrm{\mu m}$, these bunches are much longer than $\lambda_p$ and are thus suitable for experimental study of SMI. Moreover, these bunches have transverse sizes of a few tens of microns, smaller than the plasma skin depth that varies from $16$ to $50~\mathrm{\mu m}$. These bunches 
can then 
excite plasma waves in the narrow bunch limit, suppressing growth of competing current filamentation instabilities \cite{cfi}. Initially 
they 
also excite plasma waves in the linear regime because the bunch density is much smaller than the plasma density ($n_b/n_0\simeq 10^{-3}-10^{-1}$). Figure~\ref{fig:setup} shows a schematic representation of the experimental layout.

In spite of the different bunch dimensions and plasma densities used in the self-modulated proton driven plasma wakefield accelerator and in the self-modulation experiment illustrated in Fig.~\ref{fig:setup}, both share key features enabling growth of the self-modulation instability. Namely the bunch lengths are much longer than the plasma wavelength, the bunch radius is smaller than the plasma skin-depth, and bunch densities are much smaller than the plasma density. Key physics of self-modulation of the future SM-PDPWFA experiment can therefore be studied using these long electron bunches. Moreover, these scenarios can also be of relevance in astrophysics and might lead to the generation of relativistic shocks when using bunches with transverse sizes much larger than the plasma skin-depth~\cite{bib:astro}.

In order to ensure that experiments can reach SMI saturation, the hosing instability (HI) needs to be avoided. The HI, which consists in unstable oscillations of the bunch centroid, can lead to beam breakup, thereby preventing SMI growth and saturation. In order to suppress hosing after SMI saturation in linear regimes, strong SMI seeding is required~\cite{bib:vieira_prl_2014}. This can be achieved, for instance, by using bunches with short rise times that maximize the initial wakefield amplitudes. As a result, bunches with sharp rise times are important for successful SMI experiments and hence for particle acceleration in SMI driven wakefields. In this work we show that bunches with short rise times are also critical to maximize wakefield after SMI saturation in the absence of hosing. 

Self-modulation of long, SLAC FACET electron (and positron) bunches was examined in Ref.~\cite{bib:vieira_pop_2012} using sharp rise time (step function) to seed the SMI and avoid HI. Simulations showed that multi GeV energy gain and loss are reached by the drive bunch particles after a 1 meter long plasma. In addition, they showed growth and saturation of SMI over the first few cm of propagation. 
Here we show that the use of bunches with sharp rise times is important to maximize wakefield excitation even in the absence of hosing. 
To avoid hosing we use 2D cylindrically symmetric simulations, which cannot capture hosing.
In Section~\ref{sec:simulations} we present simulation results obtained with the particle in cell code OSIRIS~\cite{bib:fonseca_book}, 
an electromagnetic, relativistic, massively parallel PIC code which makes no physical approximations as long as quantum effects can be neglected. Generally, in the particle-in-cell algorithm the spacial domain is subdivided into grid cells, each containing simulation particles. Electromagnetic fields and currents are defined in the grid. Particles are pushed using a relativistic Boris pusher using the interpolated fields at the particles positions. After pushing particles, the new currents are then deposited in the grid cells, and the discrete set of Maxwell's equations is used to advance the fields. 
In Section~\ref{sec:simulations} we compare the bunch energy spectra and maximum accelerating fields using different plasma densities and bunch rise time profiles. These simulations show that bunches with finite rise times can reduce maximum SMI driven wakefields. In Section~\ref{sec:risetime} we show that bunches with up ramps smaller than $\lambda_p/2$ are required for effective wake excitation in SMI scenarios. 
In Section~\ref{sec:conclusions} we present our conclusions.

\section{\label{sec:simulations}Self-modulation instability of electron bunches with 
finite 
rise times}

We consider electron bunch with profile given by: 
\begin{eqnarray}
\label{eq:bunch-density}
n_b = n_{b0} \frac{\xi}{\sigma_{\mathrm{rise}}} \exp{\left(-\frac{r^2}{2 \sigma_r^2}\right)}, ~~~~~~\mathrm{0<\xi<\sigma_{\mathrm{rise}}} \nonumber \\
n_b = n_{b0} \exp{\left(-\frac{r^2}{2 \sigma_r^2}\right)}, ~~~~~~~~~\mathrm{\sigma_{\mathrm{rise}}<\xi<\sigma_z},
\end{eqnarray}
where $\xi=z-ct$, $z$ is the distance, and $t$ the time. In addition, $n_{b0}=2.16\times10^{15}~\mathrm{cm}^{-3}$ is the peak bunch density, $\sigma_r = 30~\mathrm{\mu m}$ is the bunch radius, $\sigma_{\mathrm{rise}}=300~\mathrm{\mu m}$ is the bunch rise time and $\sigma_z=1.5~\mathrm{mm}$ its length. %
We note that the bunch ramp profile was chosen for ease of calculation, and that the specific ramp profile does not affect the conclusion of this work. %
The bunch energy is 19.55 GeV. We performed simulations at two different plasma densities with 
$n_0^{\mathrm{(1)}}=10^{17}~\mathrm{cm}^{-3}$ and $n_0^{\mathrm{(2)}}=4.6\times10^{17}~\mathrm{cm}^{-3}$. Hence $k_p^{\mathrm{(1)}} \sigma_r = 0.28 $ and $k_p^{\mathrm{(2)}} \sigma_r = 0.58$, $\sigma_z = 14 \lambda_p^{\mathrm{(1)}}$ and $\sigma_z = 30 \lambda_p^{\mathrm{(2)}}$, and $\sigma_{\mathrm{rise}}=2.7~\lambda_p^{\mathrm{(1)}}$ and $\sigma_{\mathrm{rise}}=6.0 \lambda_p^{\mathrm{(2)}}$ for the lower and higher plasma density, respectively. In addition, the normalised bunch emittance is $\epsilon_N=30~\mathrm{mm\cdot mrad}$. Simulations used a computational box with dimensions $2\times0.4~\mathrm{mm}^2$ and $2.2\times0.85~\mathrm{mm}^2$. The box is divided into $4764\times640$ and $2483\times1369$ cells with $2\times2$ plasma and bunch particles per cell. 
Quadratic particle shapes are also used. 
These bunch and plasma parameters could be readily obtained in the laboratory.

\begin{figure}
\centering\includegraphics[width=\columnwidth]{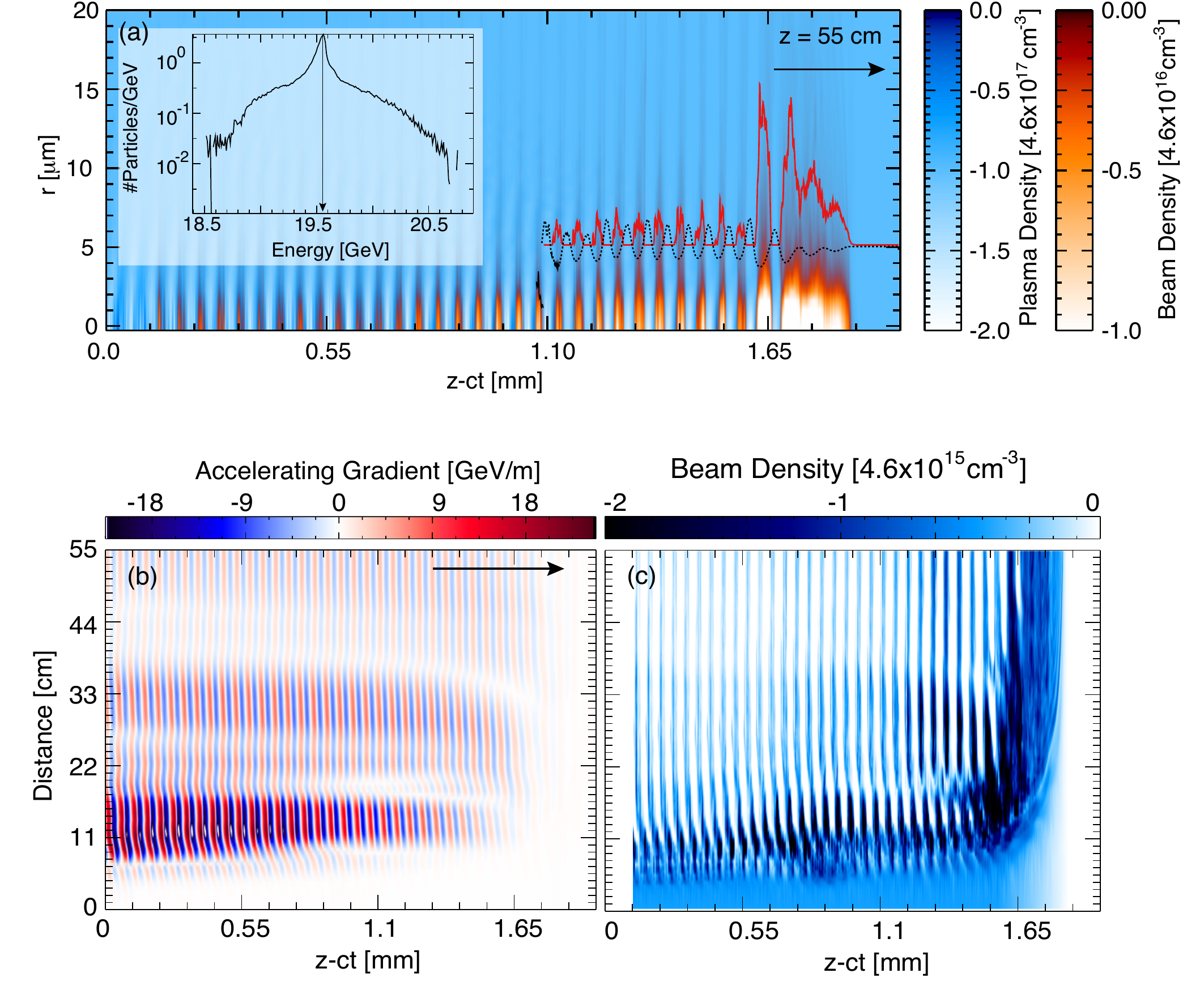}
\caption{(a) simulation result of a long electron bunch after propagation of $55~\mathrm{cm}$ in a plasma with $n_0=4.6\times 10^{17}~\mathrm{cm}^{-3}$. Plasma electron density is shown in white-blue colours and the electron bunch in white-red colours. The inset illustrates the corresponding energy spectra and shows 1 GeV energy gain and energy loss by self-modulated bunch particles. The arrow indicates the bunch propagation direction. 
Lineouts of bunch density profile (red solid line) and electron plasma density (black dotted line) near the axis are also shown. (b) and (c) show the evolution of the on-axis acceleration gradients and bunch density profile during propagation.
}
\label{fig:smi}
\end{figure}

Figure~\ref{fig:smi}a shows the self-modulated bunch after propagating 55 cm in the higher density plasma ($n_0=4.6\times10^{17}~\mathrm{cm}^{-3}$). Self-modulated beamlets propagate in regions of smaller plasma density (lighter blue) where the transverse wakefield is focusing for electrons. It is interesting to note that the front of the bunch is not fully self-modulated as it is longer than $\lambda_p$. 
This is also shown by the solid red and dashed black curves in Fig.~\ref{fig:smi}a, which represent the bunch and plasma density near the axis, respectively.
These results contrast with simulation results using electron bunches with sharp rise time in comparison to $\lambda_p$ for which all beamlets 
(including the first) 
are shorter than $\lambda_p$. Figure~\ref{fig:smi}a also shows that, unlike simulation results using bunches with sharp rise times~\cite{bib:vieira_pop_2012}, a significant fraction of the bunch is lost at the bunch second half. The inset of Fig.~\ref{fig:smi}a illustrates corresponding energy spectrum showing 1 GeV energy gain/loss after the 55 cm long plasma, much smaller than the acceleration gradients from Ref.~\cite{bib:vieira_pop_2012} which used bunches with similar charges.


The self-modulation dynamics is more complex when the bunch rise time is 
finite than when $\sigma_{\mathrm{rise}}=0$.
Figure~\ref{fig:smi}b illustrates the evolution of self-modulated driven accelerating gradients for $\sigma_{\mathrm{rise}}=2.7\lambda_p$. It shows that the wakefield amplitude has strong variations during propagation. In addition, the trajectories of the points of minimum/maximum wake amplitudes indicate strong variations of the wake phase velocity ($v_{\phi}$) from %
being %
much smaller than $c$ to regions where $v_{\phi}>c$. 
We note that if the red/blue regions in Fig.~\ref{fig:smi}b bend to the left/right then the phase velocity is lower/higher than c. 

SMI saturation at the back of the driver is seen after $\simeq 11~\mathrm{cm}$ in Fig.~\ref{fig:smi}c, which illustrates the evolution of the bunch density near the axis as a function of the propagation distance. At $z=11~\mathrm{cm}$, however, there is still little self-modulation at the bunch front, which starts to be more strongly modulated only after the bunch propagated for $22-33~\mathrm{cm}$. Wakefield phase shifts, and wake phase velocity variations are then observed even after the back of the bunch is fully self-modulated in Fig.~\ref{fig:smi}b. These wakefield variations are then due to radial modulations at the front of the bunch after SMI saturation at the back.

\begin{figure}
\centering\includegraphics[width=\columnwidth]{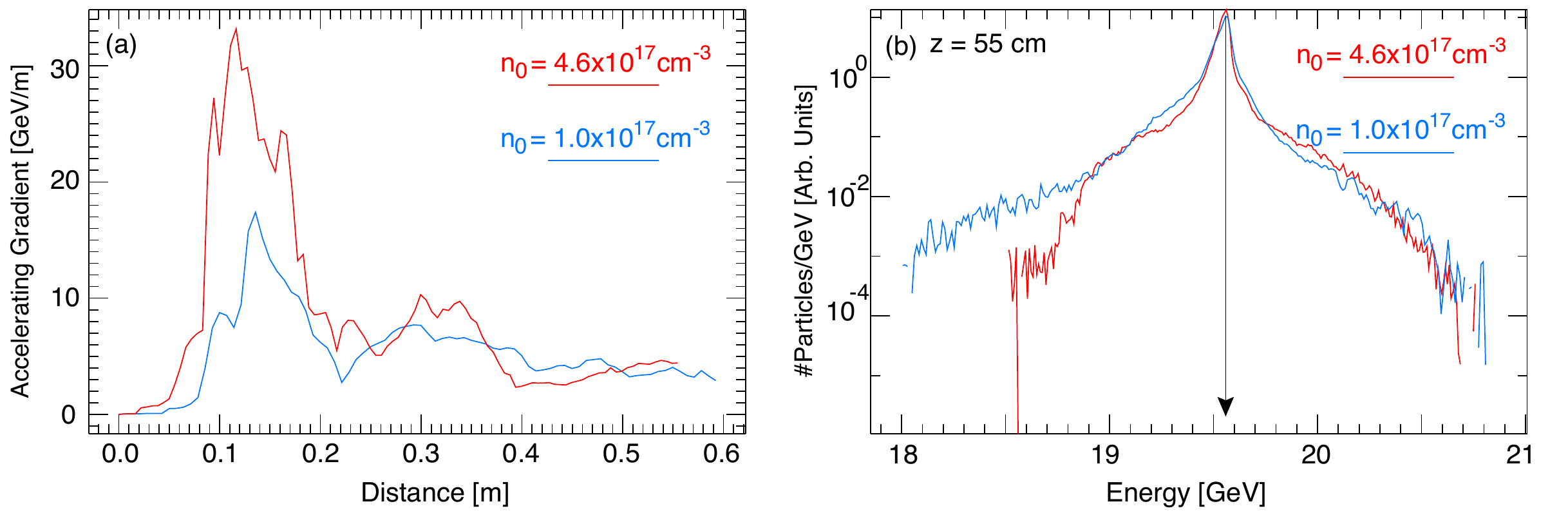}
\caption{Simulation results showing the maximum acceleration fields near the axis as a function of the propagation distance (a), and self-modulated particle bunch energy spectra after propagation of 55 cm in a plasma with density $n_0=4.6\times10^{17}~\mathrm{cm}^{-3}$ (red line) and with $n_0=1\times10^{17}~\mathrm{cm}^{-3}$. The arrow indicates the initial bunch energy in (b). }
\label{fig:comparison}
\end{figure}

Figure~\ref{fig:comparison}a compares the evolution of the maximum plasma accelerating fields measured near the axis for the two densities stated above. In agreement with Fig.~\ref{fig:smi}, Fig.~\ref{fig:comparison}a shows strong accelerating field variations. Although at higher plasma density the accelerating gradients is larger in comparison to the lower plasma density case, few GeV/m accelerating gradients are reached for both plasma densities at the end of the plasma. 
Figure~\ref{fig:comparison}a also shows that the amplitude of the wakefield rises until the bunch has propagated for $\simeq 11~\mathrm{cm}$, as the back of the bunch becomes fully self-modulated. The wakefield amplitude then drops significantly until $z\simeq30~\mathrm{cm}$, when the front of the bunch becomes more strongly modulated. These results then indicate that modulations of the bunch front after SMI saturation at the back are associated with significant reductions of maximum accelerating wakefields.
Figure~\ref{fig:comparison}b shows particle bunch energy spectra after 55 cm. The energy spectra are similar at both plasma densities, showing 1 GeV energy gain/loss in both cases.

\begin{figure}
\centering\includegraphics[width=\columnwidth]{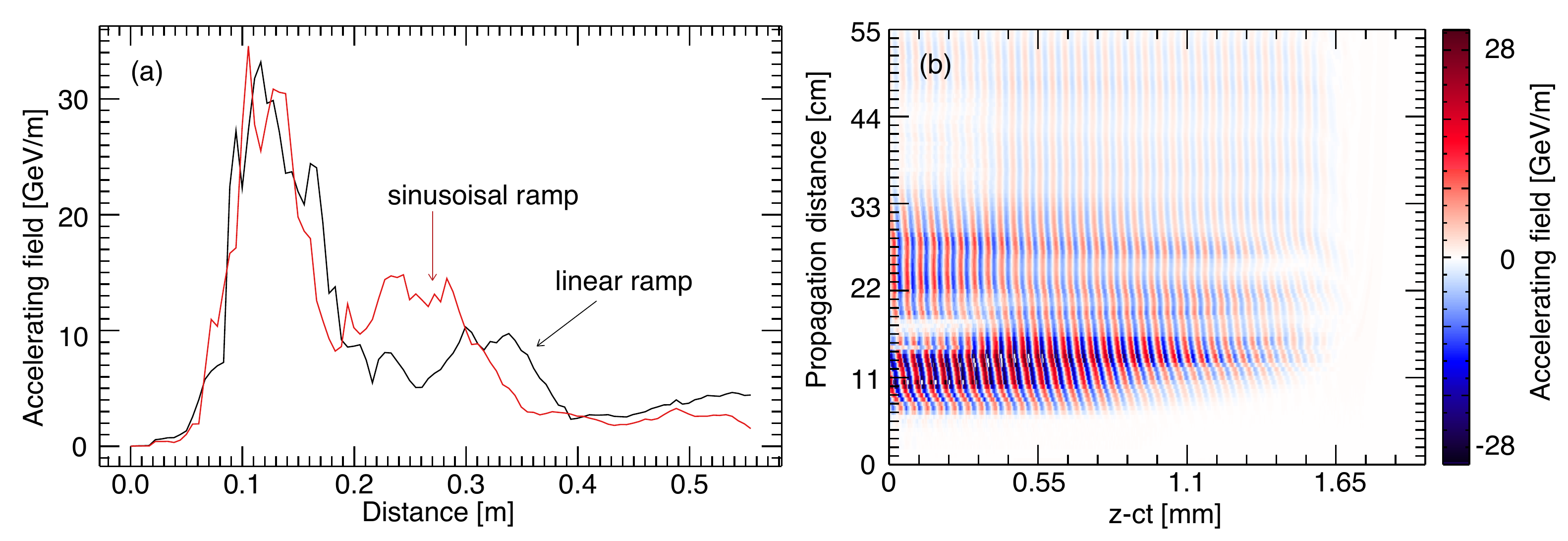}
\caption{(a) Comparison between maximum accelerating wakefields for the case of a bunch with a sinusoidal density ramp (red) and for a linear density ramp (black). (b) Evolution of the accelerating wakefields driven by a bunch with a sinusoidal ramp near the axis as a function of the propagation distance.}
\label{fig:sinusoidalramp}
\end{figure}

To confirm that the specific ramp profile does not change the above mentioned conclusions, we performed an additional simulation using a bunch profile with a sinusoidal up ramp of the form $n_b=(n_{b0}/2)\left\{\sin\left[\xi/\sigma_{\mathrm{rise}}-\pi/2\right]\right\}\exp\left(-r^2/(2 \sigma_r^2)\right)$ for $0<\xi<\sigma_{\mathrm{rise}}$ and $n_b=n_{b0}\exp\left(-r^2/(2 \sigma_r^2)\right)$ for $\sigma_{\mathrm{rise}}<\xi<\sigma_{z}$ with $\sigma_{\mathrm{rise}}=\sqrt{2}\lambda_p$. The bunch propagates in a plasma with $n_0=4.6\times10^{17}~\mathrm{cm}^{-3}$. Results, shown in Fig.~\ref{fig:sinusoidalramp} indicate that our conclusions are independent of the specific bunch ramp profile. Figure~\ref{fig:sinusoidalramp}a provides a comparison of the maximum accelerating wakefields as a function of the propagation distance between a linear and sinusoidal up-ramp density profiles. The general trend is similar for both cases. Similarly to the results shown in Fig.~\ref{fig:smi}, self-modulation also saturates faster at the bunch back than at the front when a sinusoisal ramp profile is used. Figure~\ref{fig:sinusoidalramp}b shows that strong wakefield phase and wakefield phase velocity variations occur during the self-modulation of a bunch with a sinusoidal up ramp. Figure~\ref{fig:sinusoidalramp} then confirms that our conclusions are independent of the specific bunch ramp density profile. 

\section{\label{sec:risetime}Bunch rise time requirements for maximum self-modulated wakefield amplitudes}

Results from Fig.~\ref{fig:smi}, Fig.~\ref{fig:comparison} and Fig.~\ref{fig:sinusoidalramp} are in stark contrast with scenarios where the wakefield is driven by a hard-cut bunch with $\sigma_{\mathrm{rise}}\ll \lambda_p$~\cite{bib:vieira_pop_2012}. Specifically, the wakefields amplitude is maximised and it is much more stable when $\sigma_{\mathrm{rise}}\ll \lambda_p$ than when $\sigma_{\mathrm{rise}}\gtrsim \lambda_p$. The acceleration of external particles can therefore become less effective if the bunch has a finite rise time: on the one hand, wakefield phase velocity variations can lower the dephasing length and make particle trapping more difficult. On the other hand, lower wakefield amplitudes also lead to lower final energies. 

\begin{figure}
\centering\includegraphics[width=\columnwidth]{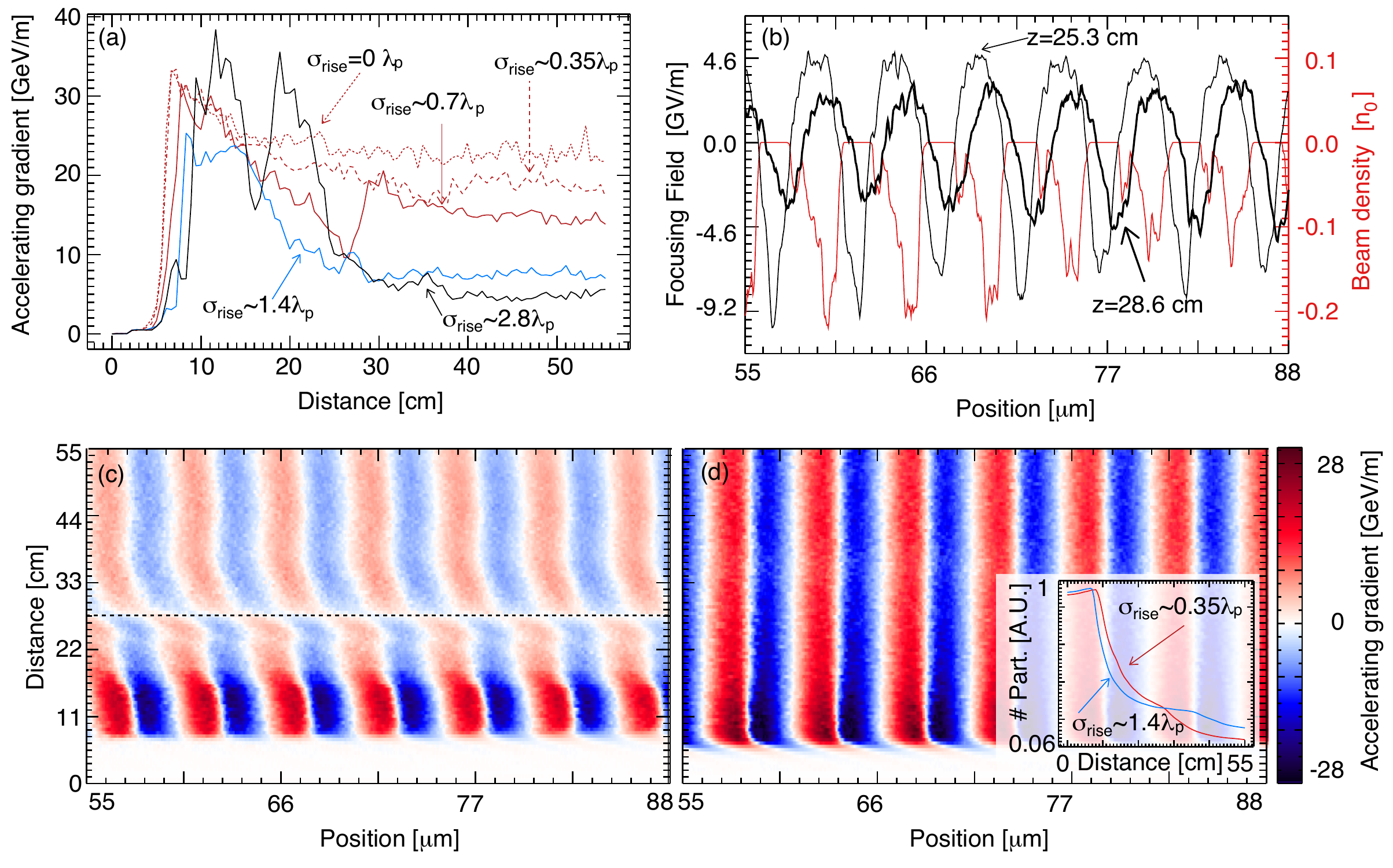}
\caption{(a) maximum accelerating gradients as a function of the propagation distance for bunches with different rise times. (b) Beam density line out (red) at $z=25~\mathrm{cm}$ and focusing fields lineouts near the axis for the bunch with $\sigma_{\mathrm{rise}} = \lambda_p$. Lineouts were taken at $z=25~\mathrm{cm}$ (black thiner line) and at $z=28~\mathrm{cm}$ (black thicker line). The tail of the bunch is at $0\mathrm{\mu m}$. (c) and (d) show the corresponding accelerating wakefield evolution for $\sigma_{\mathrm{rise}}=\lambda_p$ and $\sigma_{\mathrm{rise}}=\lambda_p/4$, respectivelly. The dashed line in (c) indicates the approximate location where the lineouts in (b) were taken. The inset in (d) shows the evolution of the number of beam electrons inside the region $r<2\sigma_r$.}
\label{fig:ramplength}
\end{figure}

In order to understand how do these conclusions change for bunches with different rise times, we performed additional numerical simulations with different $\sigma_{\mathrm{rise}}$. Simulations indicate that bunches with ramp sizes smaller than $\sigma_{\mathrm{rise}}\lesssim \lambda_p/2$ are needed to maximize the wakefield amplitudes. This can be seen in Fig.~\ref{fig:ramplength}a which illustrates the evolution of the maximum wakefields for bunches with different $\sigma_{\mathrm{rise}}$ propagating in a plasma with $n_0=4.6\times10^{17}~\mathrm{cm}^{-3}$. We performed simulations for a hard cut bunch with $\sigma_{\mathrm{rise}}=0$, and also using $\sigma_{\mathrm{rise}}=\sqrt{2}\lambda_p/4\simeq 0.35 \lambda_p$, $\sigma_{\mathrm{rise}}=\sqrt{2}\lambda_p/2\simeq 0.7\lambda_p$, $\sigma_{\mathrm{rise}}=\sqrt{2}\lambda_p\simeq 1.4\lambda_p$, and $\sigma_{\mathrm{rise}}=2\sqrt{2}\lambda_p\simeq2.8\lambda_p$. According to Fig.~\ref{fig:ramplength}a, smaller $\sigma_{\mathrm{rise}}$ enhance the maximum wakefields after SMI saturation. Wakefields are close to maximum for $\sigma_{\mathrm{rise}}\lesssim \lambda_p/2$.

These results can be attributed to the stronger phase velocity variations and stronger variations of the wakefield phase for the bunches with larger $\sigma_{\mathrm{rise}}$. Phase velocity variations can reduce the number of particles available to drive the plasma wakefields. To understand how, we show in Fig.~\ref{fig:ramplength}b the bunch density near the axis for $\sigma_{\mathrm{rise}}=\lambda_p$ at $z=25.3~\mathrm{cm}$ superimposed with the focusing fields at $z=25.3~\mathrm{cm}$ (black thinner solid line) and at $z=28.6~\mathrm{cm}$ (black thicker solid line). The wake phase velocity changes between these two propagation distances since the focusing force extrema move forward with respect to the bunch, which remains almost unchanged since it moves at c. As a result, some of the beamlet electrons, which resided in focusing field regions (i.e. positive focusing force values) at $z=25~\mathrm{cm}$, become in defocusing field regions at $z=29~\mathrm{cm}$. Simulations then show that these particles diffract radially, being lost to the plasma. This mechanism can then decrease the amount of charge available to drive the wakefields.

The wake phase velocity variations can also be deduced from Figs.~\ref{fig:ramplength}c-d for $\sigma_{\mathrm{rise}}=\sqrt{2}\lambda_p$ and $\sigma_{\mathrm{rise}}=\sqrt{2}\lambda_p/4$ respectively. Comparison between Figs.~\ref{fig:ramplength}c and Figs.~\ref{fig:ramplength}d then confirms that smaller $\sigma_{\mathrm{rise}}$ lead to smaller wake phase velocity variations. The inset of Fig.~\ref{fig:ramplength}d shows the number of bunch particles within $r<2\sigma_r$. The inset shows that the number of particles available to drive the wakefields for $\sigma_{\mathrm{rise}}=\sqrt{2}\lambda_p$ is almost 60$\%$ of those of $\sigma_{\mathrm{rise}}=\sqrt{2}\lambda_p/4$ at the end of the plasma. 

The lower number of particles available to drive the wakefields when using higher $\sigma_{\mathrm{rise}}$ contributes to the lower accelerating gradients observed in the simulations when $\sigma_{\mathrm{rise}}\gg \lambda_p/2$. However, changes in the wakefield phase due to the radial dynamics of the bunch front can also contribute to lower wakefield amplitudes by modifying the relative phase between the wakefields driven by each beamlet along the bunch. The wakefields driven by each self-modulated beamlet can then change from a scenario of constructive wakefield superposition (leading to resonant wakefield excitation), to a case of destructive wakefield superposition (which can lower wakefield amplitudes). These processes are complex and more detailed descriptions are deferred to a future publication.

\section{\label{sec:conclusions}Conclusions}

In this work we study the influence of bunches with 
finite rise times 
on the self-modulation instability of long bunches. Unlike simulation results using bunches with sharp rise time, this work shows that the evolution of the self-modulation is characterised by strong variations of the wakefield amplitude, 
wake phase and wake phase velocity until SMI saturates 
We attribute this evolution to the radial modulation variations at the bunch front, in the region of rising bunch density. These results show the importance of using bunches with sharp rise times with $\sigma_{\mathrm{rise}}\lesssim \lambda_p/2$ in order to effectively seed the SMI and to achieve a stable acceleration situation. Note that while bunches produce by accelerators or linacs may not have very sharp rise time, a masking method exists~\cite{mask} and is available at SLAC FACET~\cite{FACET} to produce bunches with rise time shorter that the plasma wavelength. 

\section{Acknowledgments}

Work partially supported by the European Research Council (ERC-2010-AdG Grant No. 267841) and by FCT (Portugal) through Grants Nos. EXPL/FIS-PLA/0834/2012. We acknowledge PRACE from awarding access to resource SuperMUC based in Germany at Leibniz research centre.


\section*{References}
\begin{thebibliography}{10}
\bibitem{bib:tajima_prl_1979} T. Tajima, J.M. Dawson, Phys. Rev. Lett. \textrm{43} 267 (1979).
\bibitem{bib:chen_prl_1986} P. Chen, J. M. Dawson, R. Huff, T. Katsouleas, Phys. Rev. Lett. \textbf{54}, 674 (1986).
\bibitem{bib:acceleration} W. P. Leemans, B. Nagler, A. Gonsalves, Cs. T\'oth, K. Nakamura, C.G.R. Geddes, E. Esarey, C.B. Schroeder, S. Hooker, Nat. Phys. \textbf{2}, 696 (2006); S. Kneip \emph{et al.}, Phys. Rev. Lett. \textbf{103}, 035002  (2009); D. H. Froula \emph{et al.}, Phys. Rev. Lett. \textbf{103}, 215006 (2009); I. Blumenfeld \emph{et al.}, Nature \textbf{445}, 741 (2007).
\bibitem{bib:blowout} A. Pukhov and J. Meyer ter Vehn, Appl. Phys. B: Lasers Opt. \textbf{74}, 355 (2002); W. Lu, C. Huang, W.B. Mori, T. Katsouleas, Phys. Rev. Lett. \textbf{96}, 165002 (2006); W. Lu, M. Tzoufras, C. Joshi, F.S. Tsung, W.B. Mori, J. Vieira, R.A. Fonseca, L.O. Silva, Phys. Rev. ST Accel. Beams \textbf{10}, 061301 (2007); 
\bibitem{bib:pdpwfa} A. Caldwell, K. Lotov, A. Pukhov, F. Simon, Nat. Phys. \emph{5}, 363 (2009).
\bibitem{awake} R. Assmann, R. Bingham, T. Bohl, C. Bracco, B. Buttenschon, A. Butterworth, A. Caldwell, S. Chattopadhyay, S. Cipiccia, E. Feldbaumer, R.A. Fonseca, B. Goddard, M. Gross, O. Grulke, E. Gschwendtner, J. Holloway, C. Huang, D. Jaroszynski, S. Jolly, P. Kempkes, N. Lopes, K. Lotov, J. Machacek, S.R. Mandry, M. Meddahi, N. Moschuering, P. Muggli, Z. Najmudin, P. A. Norreys, E. Oz, A. Pardons, A. Petrenko, A. Pukhov, K. Rieger, O. Reimann, H. Ruhl, E. Shaposhnikova, L.O. Silva, A. Sosedkin, R. Tarkeshian, R.M.G.N. Trines, T. Tuckmantel, J. Vieira, H. Vincke, M. Wing, G. Xia, this issue.
\bibitem{bib:suckin} S. Lee, T. Katsouleas, R.G. Hemker, E.S. Dodd, W.B. Mori, Phys. Rev. E \emph{64}, 045501(R) (2001). 
\bibitem{cfi} B. Allen, V. Yakimenko, M. Babzien, M. Fedurin, K. Kusche, P. Muggli, Phys. Rev. Lett. 109, 185007 (2012).
\bibitem{bib:smi} N.Kumar, A.Pukhov,and K.Lotov, Phys. Rev. Lett. \emph{104}, 255003 (2010).
\bibitem{bib:vieira_pop_2012} J. Vieira, Y. Fang, W.B. Mori, L.O. Silva, P. Muggli, Phys. Plasmas \textbf{19}, 063105 (2012). 
\bibitem{bib:astro} L.O. Silva, AIP Conf. Proc. \textbf{856}, 109 (2006).
\bibitem{bib:vieira_prl_2014} J. Vieira, W.B. Mori, P. Muggli, accepted for publication in Phys. Rev. Lett. (2014).
\bibitem{bib:fonseca_book}R. A. Fonseca \emph{et al.}, Lect. Notes Comp. Sci. vol. 2331/2002, (Springer Berlin / Heidelberg,(2002); R. A. Fonseca, S. F. Martins, L. O . Silva, F. S. Tsung, W. B. Mori, Plasma Phys. Control. Fusion \textrm{50} 124034 (2008).
\bibitem{mask} P. Muggli, V. Yakimenko, M. Babzien, E. Kallos, K. P. Kusche, Phys. Rev. Lett. 101, 054801 (2008), Y. Fang, V. E. Yakimenko, M. Babzien, M. Fedurin, K. P. Kusche, R. Malone, J. Vieira, W.B. Mori and P. Muggli, to appear in Phys. Rev. Lett.
\bibitem{FACET} M.J. Hogan, T. O. Raubenheimer, A. Seryi, P. Muggli, T. Katsouleas, C. Huang, W. Lu, W. An, K.A. Marsh, W.B. Mori, C. E. Clayton, C. Joshi, New J. Phys. 12, 055030 (2010).
\end{thebibliography}
\end{document}